# Photoacoustic elastic oscillation and characterization

Fei Gao[1,*], Xiaohua Feng[1], and Yuanjin Zheng[1]

[1]*School of Electrical and Electronic Engineering, Nanyang Technological University, 639798, Singapore*

[*]Corresponding author: fgao1@e.ntu.edu.sg

**Abstract:**

Photoacoustic imaging and sensing have been studied extensively to probe the optical absorption of biological tissue in multiple scales ranging from large organs to small molecules. However, its elastic oscillation characterization is rarely studied and has been an untapped area to be explored. In literature, photoacoustic signal induced by pulsed laser is commonly modelled as a bipolar "N-shape" pulse from an optical absorber. In this paper, the photoacoustic damped oscillation is predicted and modelled by an equivalent mass-spring system by treating the optical absorber as an elastic oscillator. The photoacoustic simulation incorporating the proposed oscillation model shows better agreement with the measured signal from an elastic phantom, than conventional photoacoustic simulation model. More interestingly, the photoacoustic damping oscillation effect could potentially be a useful characterization approach to evaluate biological tissue's mechanical properties in terms of relaxation time, peak number and ratio beyond optical absorption only, which is experimentally demonstrated in this paper.

## 1. Introduction

Photoacoustic (PA) effect refers to the ultrasound generation induced by the pulsed laser illumination due to thermoelastic expansion [1]-[2]. Based on the "listening to photons" advantage of PA effect to break through the optical diffusion limit, both PA microscopy and computed tomography have attracted dramatically increasing research interest in recent years to achieve high optical contrast and high ultrasound resolution at unprecedented imaging depth [3]-[11]. Theoretically, to better understand and model the PA generation mechanism, both analytical and numerical expressions to simulate the PA generation and propagation have been derived in previous literatures [12]. The simulated PA waveform from existing analytical model or numerical simulator (i.e. k-space pseudospectral method [13]-[14]) is commonly a bipolar "N-shape" pulse with no oscillation effect. However in real situation, the PA oscillation effect could be considered as the damped harmonic oscillation of an elastic optical absorber, which has not been properly modelled yet. In this paper, firstly we derive the PA generation equation and model it as a damped mass-spring system. Then the PA damped oscillation effect is predicted and simulated as an impulse response of the damped mass- spring oscillator. The simulated PA signal shows much better coincidence with the measured PA signal compared with the ones by existing k-space pseudospectral method. Moreover, the PA damped oscillation could potentially be a very useful tool to characterize the mechanical properties, e.g. viscoelasticity and acoustic impedance, of the object in terms of the oscillation relaxation time, peak number and ratio. A feasibility study using two types of phantoms (a black line made of plasticized polyvinyl chloride, and a silicone tube filled with porcine blood) is demonstrated to show different PA oscillation effect. Ex vivo porcine muscle tissues (pink muscle and red muscle) are also evaluated and differentiated by their oscillation peak ratios. Lastly, a simulation study of image reconstruction embedding the PA damped oscillation effect is provided to demonstrate more details for tumor characterization than traditional PA imaging.

## 2. Theory of photoacoustic elastic oscillation

The formulation of PA effect has been well established in previous literatures [1-2, 12]. By modelling the optical absorbing object as a viscid point source shown in Fig. 1(a), the generated PA wave will follow the below equation [15-16]:

$$\frac{\partial^2}{\partial t^2} p(t) + a^2 \frac{\xi + \frac{4}{3}\eta}{\rho} \frac{\partial}{\partial t} p(t) + a^2 c^2 p(t) = \Gamma \frac{\partial H(t)}{\partial t}, \quad (1)$$

where $a$ is the propagation phase constant, $\rho$ is the tissue density, $\eta$ is the shear viscosity, $\xi$ is the bulk viscosity, and $c$ is the acoustic velocity in tissue. $\Gamma$ is the Gruneisen constant expressed as $\Gamma = \beta c^2 / c_p$, where $\beta$ is the thermal expansion coefficient, and $c_p$ is the constant pressure heat capacity per unit mass. $H(t) = \mu_a \Phi(t)$ is the heating function of laser illumination, where $\mu_a$ is the optical absorption coefficient of tissue, and $\Phi(t)$ is the optical radiation fluence rate.

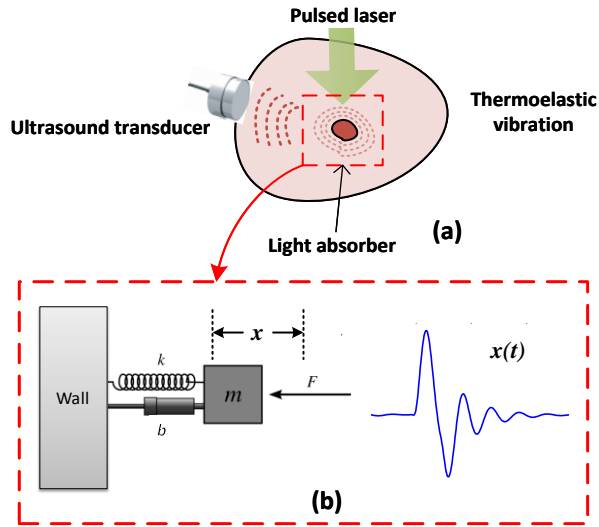

Fig. 1. (a) The diagram of PA effect induced by pulsed laser. (b) The mass-spring damped oscillation model of PA effect and a typical simulated PA signal incorporating this model.

It is obvious that Eq. (1) is a second order differential pressure equation driven by the optical source term $\Gamma \times \partial H(t)/\partial t$. Therefore, in order to give an intuitive perspective of the PA oscillation, we employ the well-known damped mass-spring system for illustration [17], as shown in Fig. 1(b). The differential equation of displacement $x(t)$ of this oscillator is:

$$\frac{\partial^2}{\partial t^2}x(t) + \frac{b}{m}\frac{\partial}{\partial t}x(t) + \frac{k}{m}x(t) = \frac{F(t)}{m}, \qquad (2)$$

where $m$ is the mass, $k$ is the string constant, and $b$ is the damping coefficient, which represents a force proportional to the speed of the mass. Comparing Eq. (1) and Eq. (2), it is observed that they agree well as a second order differential equation with the similar source term. Therefore, we could naturally map the parameters in Eq. (1) and Eq. (2) to obtain:

$$\begin{aligned}\frac{1}{m} &= \Gamma, \\ \frac{b}{m} &= a^2 \frac{\xi + \frac{4}{3}\eta}{\rho}, \\ \frac{k}{m} &= a^2 c^2.\end{aligned} \qquad (3)$$

To derive the analytical solution of Eq. (2), the PA response of the elastic object is treated as an impulse response externally stimulated by an ultra-short laser pulse. In this case the source term $F(t)$ could be simplified to be zero when the laser pulse-width (~ 2 ns) is much shorter than the stress relaxation time (~1.3 μs). Then the laser excitation is in stress confinement and could be treated as impulse excitation. By substituting $\omega_0^2 = k/m, \zeta = b/2m$ into Eq. (2), we can have:

$$\frac{\partial^2}{\partial t^2}x(t) + 2\zeta \frac{\partial}{\partial t}x(t) + \omega_0^2 x(t) = 0, \qquad (4)$$

then it could be solved by:

$$x(t) = Ae^{s_1 t} + Be^{s_2 t},$$
$$s_1 = -\zeta + \sqrt{\zeta^2 - \omega_0^2},$$
$$s_2 = -\zeta - \sqrt{\zeta^2 - \omega_0^2},$$
(5)

where $A, B$ could be determined by the initial conditions. The solution in Eq. (5) could be further categorized into three cases:

1. $\zeta < \omega_0$: under damped case:

$$x(t) = A_d e^{-\zeta t} \cos(\omega_d t + \phi_d), \text{ where } \omega_d = \sqrt{\omega_0^2 - \zeta^2}. \quad (6)$$

The constant $A_d, \phi_d$ are to be determined by the initial conditions. The motion $x(t)$ is an exponentially damped sinusoidal oscillation.

2. $\zeta > \omega_0$: over damped case:

$$x(t) = A_1 e^{-\lambda_1 t} + A_2 e^{-\lambda_2 t},$$
$$\lambda_1 = \zeta + \sqrt{\zeta^2 - \omega_0^2},$$
$$\lambda_2 = \zeta - \sqrt{\zeta^2 - \omega_0^2}.$$
(7)

The constant $A_1, A_2$ are to be determined by the initial conditions. There is no oscillation at all and the motion $x(t)$ dies off exponentially.

3. $\zeta = \omega_0$: critically damped case:

$$x(t) = A_c e^{-\zeta t}. \quad (8)$$

The constant $A_c$ is to be determined by the initial conditions. This is the case where motion $x(t)$ dies off in the quickest way with no oscillation.

Among the above three cases, because PA effect includes both expansion and contraction caused by transient laser-induced heating and sequent cooling, the generated PA signal is normally following the under damped oscillation including both positive and negative peaks, as shown in Fig. 1(b). By substituting the parameters in Eq. (3) into Eq. (6), we can have:

$$p(t) = A_d e^{-\frac{1}{2}a^2 \frac{\xi + \frac{4}{3}\eta}{\rho} t} \cos\left(\sqrt{a^2 c^2 - \left(\frac{1}{2}a^2 \frac{\xi + \frac{4}{3}\eta}{\rho}\right)^2} \, t - \frac{\pi}{2}\right), \text{ when } p(0) = 0. \tag{9}$$

According to Eq. (9), the derived PA oscillation signal is a multiplication of a sinusoidal function and a negative exponential function. Therefore, the PA oscillation signal could be treated as a sinusoidal signal with its amplitude attenuated exponentially over time. To visualize the properties of PA damped oscillation for object characterization, here three parameters are defined: relaxation time $T_r$, peak number $P_n$ and peak ratio $P_r$. Relaxation time is defined as the time duration when the envelope amplitude of the PA signal is decreased to 10% of the first maximum peak:

$$e^{-\frac{1}{2}a^2 \frac{\xi + \frac{4}{3}\eta}{\rho} T_r} = 0.1. \tag{10}$$

Then the relaxation time could be obtained by:

$$T_r = \frac{\rho}{a^2} \frac{4.6}{\xi + \frac{4}{3}\eta}. \tag{11}$$

It is observed from Eq. (11) that the relaxation time $T_r$ is proportional to the density $\rho$, and inversely proportional to the composite viscosity $\xi + 4/3\eta$. The peak number $P_n$ could be derived by the ratio between relaxation time $T_r$ and the sinusoidal period $T = 2\pi/\omega_d$:

$$P_n = \frac{T_r}{T} = \frac{2.3}{2\pi} \sqrt{\frac{4}{a^2}\left(\frac{\rho c}{\xi + \frac{4}{3}\eta}\right)^2 - 1}. \tag{12}$$

It is observed from Eq. (12) that the peak number $P_n$ is proportional to the ratio between acoustic impedance $\rho c$ and viscosity $\xi + 4/3\eta$, which could be a novel characterization parameter correlating optical absorption, mechanical property (bulk and shear viscosity) and acoustic property (acoustic impedance) of the same object. Lastly, for some extreme cases, where the samples are generating similar PA waveforms, they cannot be differentiated by their relaxation time or peak number, the peak ratio $P_r$ between the first peak and second peak of PA signal could be derived for more accurate evaluation:

$$P_r = \frac{P\left(t = \dfrac{\pi}{2\omega_d}\right)}{P\left(t = \dfrac{5\pi}{2\omega_d}\right)} = \exp\left(\frac{2\pi}{\sqrt{\dfrac{4}{a^2}\left(\dfrac{\rho c}{\xi + \dfrac{4}{3}\eta}\right)^2 - 1}}\right). \tag{13}$$

Different than the peak number expression in Eq. (12), the peak ratio $P_r$ of Eq. (13) is exponentially and inversely proportional to the ratio between acoustic impedance and viscosity. Intuitively, a higher viscosity will cause larger peak ratio, indicating more energy loss per oscillation cycle. The exponential function will also enhance the sensitivity of peak ratio characterization. We will demonstrate in the following experiment that using peak ratio $P_r$ to differentiate the tissues more accurately, which even have the same peak number $P_n$.

## 3. Results and discussion

In the above section, the PA damped elastic oscillation has been well modelled by a mass-spring system. To incorporate the PA damped oscillation effect into a PA simulation tool, the k-space pseudospectral method based MATLAB toolbox was employed and modified [13]. The initial pressure was replaced by a time-varying pressure source from Eq. (9) by assigning proper physical parameters. The key parameters of the MATLAB simulation model is summarised in Table 1.

Table 1 Key parameters of simulation model of PA elastic oscillation

| *Simulation parameters* | *Value* | *Unit* |
|---|---|---|

| *Dimension* | 50×50 | mm |
| *Space step* | 0.1 | mm |
| *Time step* | 20 | ns |
| *Source diameter* | 2 | mm |
| *Acoustic velocity* | 1500 | m/s |
| *Attenuation* | 0.75 | dB/(MHz cm) |
| *Central frequency* | 1 | MHz |
| *Bandwidth* | 60 | % |

Experiment on phantom was also conducted based on the PA measurement setup, as shown in Fig. 2. A nanosecond pulsed laser (FDSS 532-1000, CryLaS, GmbH) was used to provide green light illumination with 532 nm wavelength and 1 mJ pulse energy. The light was focused into a multimode fibre (MHP550L02, Thorlabs) by a fibre coupler, and weakly focused on the phantom by a pair of lens (LB1471, Thorlabs) with a spot size of 2 mm diameter. After light coupling and fibre delivery, the energy illuminated on the sample is around 0.2 mJ, leading to 5 mJ/cm$^2$ energy density well within the laser safety limit according to ANSI standard. The black line phantom made of plasticized polyvinyl chloride (PVC) was immersed in water for optimum light transparency and acoustic coupling. A focused ultrasound transducer (V303-SU, Olympus) with 1 MHz central frequency was used to detect the PA signal, followed by a 54 dB gain preamplifier (5662, Olympus). A digital oscilloscope (WaveRunner 640Zi, LeCroy) with 500 MHz sampling rate was used to record the PA signal, which was sent to a PC for post-processing. The size and optical absorption properties of the experimental samples used are summarised in table 2.

Table 2 Size and optical absorption properties of samples

| *Experimental sample* | *Size/Diameter (mm)* | *Optical absorption (cm$^{-1}$)* |
| --- | --- | --- |
| PVC line | 3 | 618.5 |
| Silicone tube with blood | 3 | 216 |
| Pink muscle | 10 | 1.7 |
| Red muscle | 10 | 3.9 |

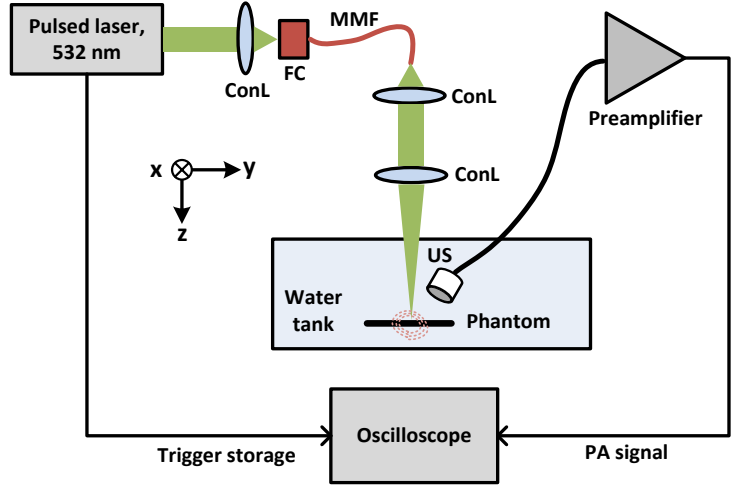

Fig. 2. The experimental setup of photoacoustic measurement. ConL: condenser lens; FC: fibre coupler; MMF: multi-mode fibre; US: ultrasound transducer.

The PA simulation results of k-space pseudospectral method without and with incorporating the PA oscillation model are shown in Fig. 3(a) and Fig. 3(b), as well as the measured PA signal shown in Fig. 3(c). It is clearly observed that by incorporating the PA oscillation model in the k-space simulation tool, the simulated PA damped oscillation waveform in Fig. 3(b) is much more similar with the measured waveform in Fig. 3(c) in shape, compared with the conventional k-space simulated waveform in Fig. 3(a). Quantitatively, the relaxation times $T_r$ of the PA waveforms in Fig. 3(a)-(c) are calculated to be 2 μs, 4.3 μs, and 4.5 μs. The peak numbers $P_n$ are 1, 4, and 4, showing that the PA simulation with the proposed oscillation model presents much better agreement than the conventional simulation approach. Fig. 3(d)-(f) are plotted to show their respective spectrums, and the quality factors match well between Fig. 3(e) and Fig. 3(f), validating the proposed PA damped oscillation model.

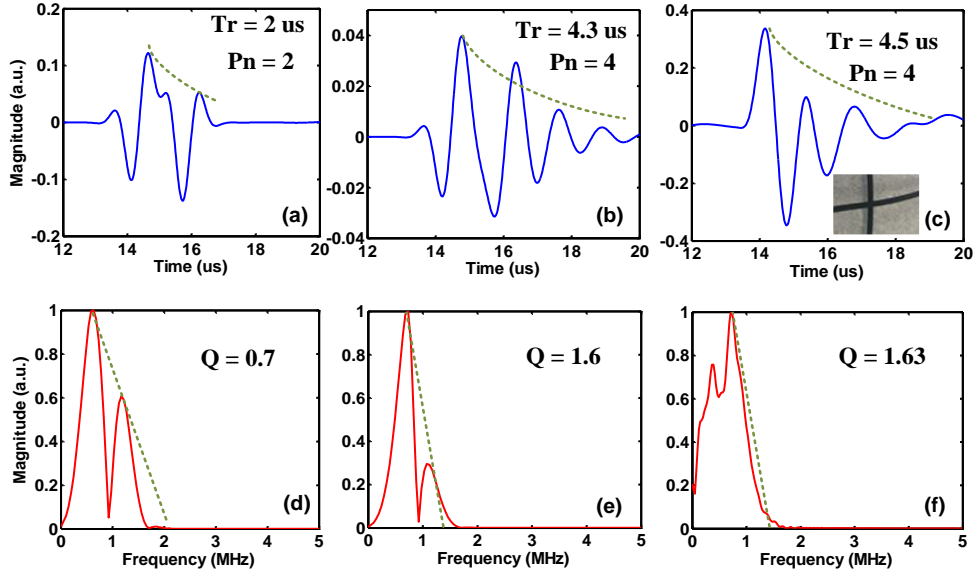

Fig. 3. (a)-(b) K-space pseudospectral method simulation results without and with incorporating the proposed PA oscillation model. (c) Measured PA signal of the black line phantom with photograph. (d)-(f) The frequency spectrums of the PA signals in (a)-(c).

To further verify the characterization capability of PA damped oscillation in terms of relaxation time and peak number, a vessel-mimicking phantom is prepared using a silicone tube filled with porcine blood. The measured PA signal is shown in Fig. 4(a), where the relaxation time is 1.2 μs and the peak number is 3. Compared with the PVC line phantom in last experiment, the vessel-mimicking phantom suffers higher viscosity and lower acoustic velocity, leading to smaller relaxation time and peak number, predicted from Eq. (11) & (12). To demonstrate the feasibility of characterizing biological tissues, ex vivo porcine tissues are prepared including pink muscle and red muscle. Fig. 4(b) and Fig. 4(c) are showing the typical PA oscillation signals from pink muscle and red muscle parts, where the peak number is same for both and invalid for characterization. Therefore we calculate the peak ratio $P_r$ of the two kinds of muscles to be 3.5 and 9.2. Then we conduct 20 measurements for both pink and red muscles at different spots within the red circle region in the tissue photograph, which are plotted in Fig. 4(d). It is observed that red muscle has statistically much higher peak ratio $P_r=8.82\pm1.28$ than pink muscle $P_r=3.64\pm0.55$, indicating the higher acoustic energy loss rate in red muscle. The underlying reason is: the red muscle contains more mitochondria, myoglobin, and capillaries than the pink muscle [18]. Therefore the red muscle is expected to suffer higher viscosity to attenuate the acoustic energy more significantly.

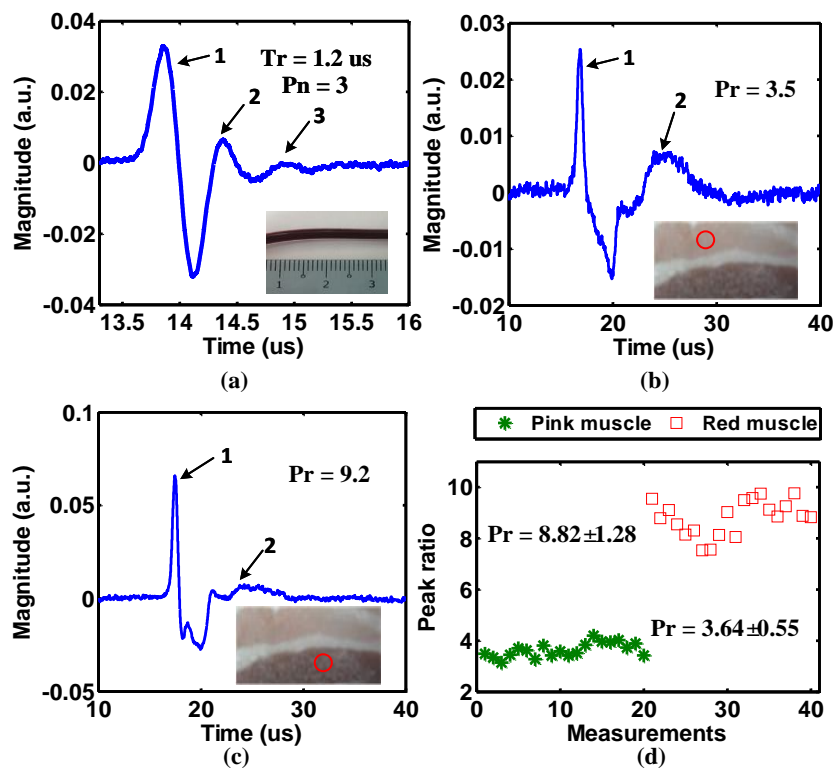

Fig. 4. (a) The typical PA oscillation signal and characterization of a vessel-mimicking phantom, (b) pink muscle, and (c) red muscle. (d) The peak ratio of 20 measurements for pink and red muscle each.

In the next section, PA imaging simulation is performed by incorporating the PA oscillation model. The simulation diagram is shown in Fig. 5(a), where two tumors are embedded with different oscillation damping factors due to different mechanical properties to mimic malignant and benign tumors, and surrounded by 100 acoustic sensors. The PA signals of the two tumors detected by one of the sensors are shown in Fig. 5(b) as well as all the sensor data shown in Fig. 5(c). It is seen that tumor 1 features larger relaxation time and peak numbers than tumor 2. After image reconstruction by simplified back-projection algorithm, it shows that the tumor 1 demonstrates more "ripples" than tumor 2 in Fig. 5(d), which indicates that tumor 1 suffers less acoustic attenuation. Based on the fact that malignant tumor has more irregular shape, greater impedance discontinuities and disorganized vasculature [19], it leads to higher viscosity and acoustic attenuation than benign tumor. Therefore, the malignant tumor will show less 'ripple' in the reconstructed images, so that we can conclude that tumor 1 is a benign tumor, and tumor 2 is a malignant one. The quantitative tumor characterization from a typical PA signal in Fig. 5(b) is provided in Table 3 in terms of relaxation time, peak number and peak ratio, providing more accurate and reliable classification of benign/malignant tumors than direct observation from the images. In this case, conventional PA imaging analysis could not tell which one is more malignant as they show similar image intensity. Interestingly, by considering the PA oscillation effect, i.e. multiple cycles in PA signal could lead to multiple ripples in PA image, the tumor more close to malignant with less ripples could be easily identified due to its distinct acoustic attenuation caused by different mechanical properties. The proposed quantitative characterization of PA elastic oscillation is verified to be potential for benign/malignant tumor classification with higher sensitivity and reliability.

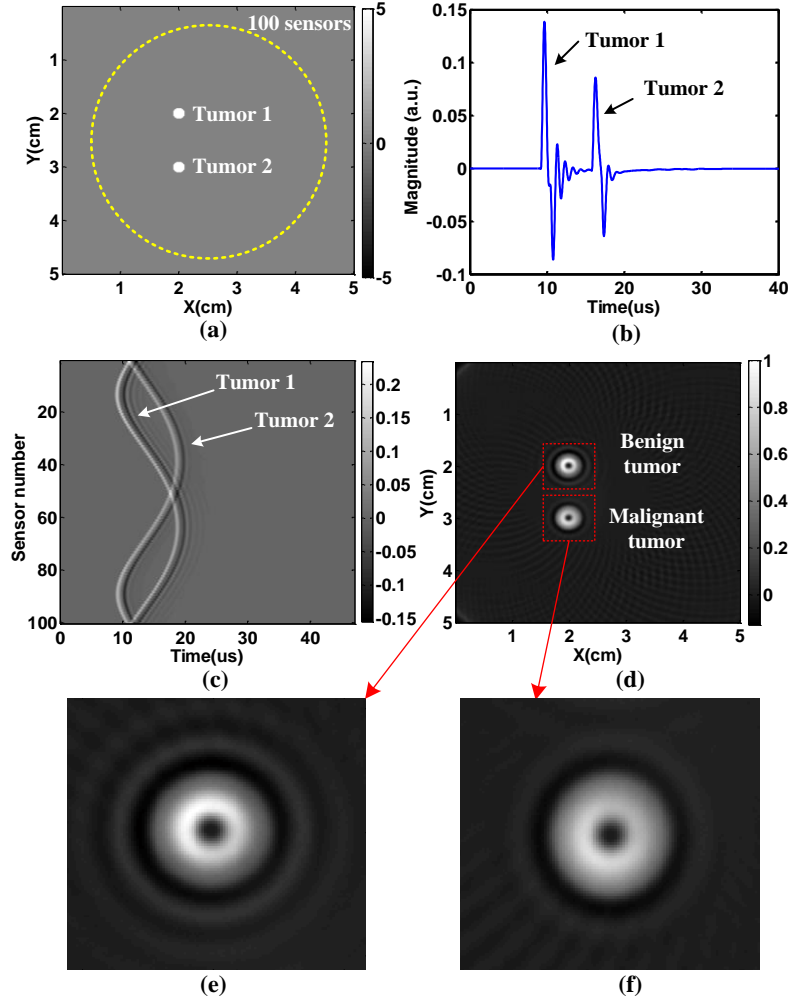

Fig. 5. (a) The PA elastic imaging simulation diagram of two tumors incorporating the PA oscillation model. (b) A typical PA signal from one of the acoustic sensors, (c) all the PA signals from 100 acoustic sensors, (d) The reconstructed PA image of the two tumors, and (e)-(f) their zoom-in figures.

Table 3 Quantitative characterization of benign/malignant tumors

|  | *Tumor 1 (benign)* | *Tumor 2 (malignant)* |
| --- | --- | --- |
| *Relaxation time ($T_r$)* | 5.6 μs | 3.5 μs |
| *Peak number ($P_n$)* | 5 | 2 |
| *Peak ratio ($P_r$)* | 6.2 | 30.7 |

Finally, the PA oscillation imaging simulation is further conducted by employing the heterogeneous numerical background to model the biologically relevant conditions. Instead of setting acoustic velocity as constant 1500 m/s, the acoustic velocity randomly varies +/- 10% (1350-1650 m/s) in the worst case with normal distribution to simulate the acoustic heterogeneity in the real soft biological tissue background, as shown in Fig. 6(a) [20-21]. The PA oscillation signal from one of the sensors is shown in Fig. 6(b), as well as all the sensor data shown in Fig. 6(c). Due to the acoustic distortion in the heterogeneous medium, the PA oscillation signals from both tumor 1 and tumor 2 are severely reshaped and degraded compared with the PA damping oscillation signal in Fig. 5(b). Even though the relaxation time ($T_r$) and peak number ($P_n$) become invalid to characterize the two different tumor types, the peak ratio ($P_r$) is still able for characterization: $P_{r1}:P_{r2}$=7.8:10.4. Therefore, it is potential that even in highly heterogeneous tissue condition, the proposed PA elastic oscillation is still possible to differentiate different types of tissues/tumors by characterizing its damping effect. As shown in Fig. 6(b) and (d), the signal patterns and reconstructed images of tumor 1 are showing more ripples than tumor 2 due to their different mechanical properties.

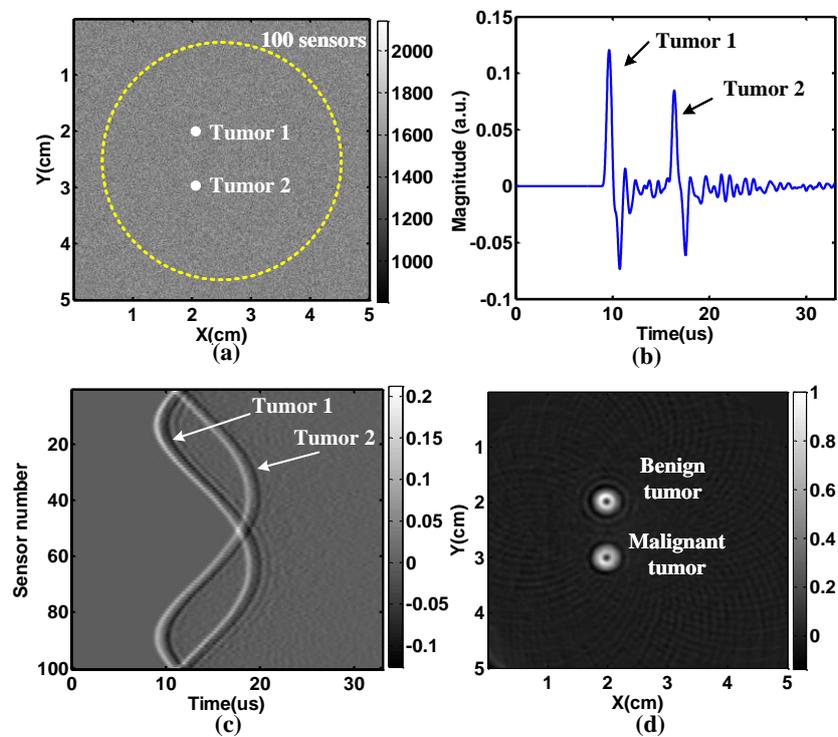

Fig. 6. (a) The PA imaging simulation diagram of two tumors incorporating the PA oscillation model with heterogeneous acoustic velocity. (b) A typical PA signal from one of the acoustic sensors, (c) all the PA signals from 100 acoustic sensors, (d) The reconstructed PA image of the two tumors.

The impact of the heterogeneous tissues on PA waveform is equivalent to the unknown impulse response of the acoustic channel. The feasibility of applying the proposed method to a tissue-mimicking condition is studied in Fig. 6 above by more complex and real-case simulations. By keeping the unknown heterogeneous map constant during the measurement, and applying advanced acoustic aberration algorithm [22], the PA oscillation characterization is also potential in a more biologically relevant condition. Regarding the detection angle and distance of the ultrasound transducer and other factors that may affect the PA waveform, we could keep transducer and the system configuration as stable as possible during the tissue characterization and imaging, where the different tissues or imaging focal point will suffer the same distortion caused by the well-fixed transducer detection distance and angle. Regarding the time-dependent variation and fluctuation (e.g. near field fluctuation) induced additional noise, one simple way is to increase data averaging to minimize the fluctuation. By doing so, potentially the parameters of the PA signal oscillation could also be extracted from the distorted detected PA signal. These factors distorting the PA signals commonly exist in all the PA imaging modalities, which could be mitigated by dedicated system design and optimization.

The frequency response of the ultrasound transducer could distort the initial PA signal waveform due to its limited bandwidth, which will induce some challenges to the PA oscillation characterization as shown in Fig. 3. However, the transducer's effect could be easily calibrated by de-convoluting its known impulse response. Even for the unknown frequency response of the transducer, it is still feasible to characterize different tissues by using the same transducer, because different initial PA signals from different kinds of tissues will experience the unknown but same frequency response of the transducer, generating distorted but still different PA waveforms for tissue characterization. To minimize the influence of the limited bandwidth of ultrasound transducer, a straightforward way is to use a broadband transducer with flat frequency response. If the dominant frequency of the PA damped oscillation signal is known, a careful frequency and bandwidth selection of transducer is preferred to cover the majority of the PA damped oscillation signal spectrum.

In this work, the optical illumination is limited to a point source to minimize the influence of the illumination and target geometry uncertainty. Therefore, the proposed PA oscillation approach is primarily potential in photoacoustic microscopy with minimum illumination and target geometry variations due to the tight optical focusing on superficial object. It immediately enables another dimension contrast of mechanical property to generate dual-contrast images, without introducing any extra components and cost. For photoacoustic tomography applications, where the PA signal profile is affected by the illumination and target geometry, as well as tissue heterogeneity mentioned above. It will be much more challenging as the PA signal may be severely distorted by these uncertain factors. Dedicated optic-acoustic confocal probe with large numerical aperture is required to minimize the geometry variation by focusing both light and ultrasound tightly in the biological tissue [10]. Advanced algorithms need to be developed to decouple the influence of acoustic heterogeneity and geometry uncertainty, which will be studied in future work.

It is worth noting that the optical absorption coefficient dominantly determines the PA signal magnitude (the first positive and negative peak-to-peak amplitude). This could introduce fluence variations in some applications, such as absolute oxygen saturation measurement [23]. Its slight influence on PA signal profile is due to the different illumination geometries by different optical decay, i.e. stronger optical delay in biological tissue, shallower illumination geometry. Dedicated optic-acoustic confocal probe with large numerical aperture is required to minimize the optical decay induced geometry variation by focusing both light and ultrasound tightly in the biological tissue. This kind of PA profile variation could be eliminated by compensating the optical absorption coefficients, which could be estimated from the PA signal amplitude.

## 4. Conclusion

In conclusion, we have modelled the PA elastic oscillation as a damped mass-spring system theoretically, and achieved prediction and simulation much closer to the experimentally observed PA signal, compared with conventional PA simulation. Utilizing the PA oscillation effect, three parameters are proposed to characterize the phantoms and biological tissues, which revealed the mechanical properties (e.g. viscosity, acoustic attenuation, and acoustic impedance) experimentally in terms of relaxation time, peak number, and peak ratio. Lastly the feasibility of PA elastic oscillation in imaging reconstruction is explored, where malignant and benign tumors are well differentiated. By incorporating the PA elastic oscillation characterization to conventional PA imaging, we create the potential to enable the PA imaging to characterize both optical and mechanical properties in single imaging modality.

**Acknowledgments**

This research is supported by the Singapore National Research Foundation under its Exploratory/Developmental Grant (NMRC/EDG/1062/2012) and administered by the Singapore Ministry of Health's National Medical Research Council.